\documentclass[prl,aps,twocolumn]{revtex4-1}%,%onecolumn,%groupedaddress]{revtex4}
\usepackage{epsfig}
\usepackage{amsfonts}
\usepackage{rotating}
\usepackage{sparticles}
\usepackage{dcolumn}
\usepackage{bm}
\usepackage{textcomp}
\usepackage{hyperref}
\hypersetup{
    bookmarks=true,         % show bookmarks bar?
    unicode=false,          % non-Latin characters in Acrobat’s bookmarks
    pdftoolbar=true,        % show Acrobat’s toolbar?
    pdfmenubar=true,        % show Acrobat’s menu?
    pdffitwindow=false,     % window fit to page when opened
    pdfstartview={FitH},    % fits the width of the page to the window
    pdftitle={My title},    % title
    pdfauthor={Author},     % author
    pdfsubject={Subject},   % subject of the document
    pdfcreator={Creator},   % creator of the document
    pdfproducer={Producer}, % producer of the document
    pdfkeywords={keyword1} {key2} {key3}, % list of keywords
    pdfnewwindow=true,      % links in new window
    colorlinks=true,       % false: boxed links; true: colored links
    linkcolor=red,          % color of internal links
    citecolor=red,        % color of links to bibliography
    filecolor=magenta,      % color of file links
    urlcolor=green,           % color of external links
    linktocpage=true
}
\usepackage{amsmath,amssymb,amsthm,graphicx,latexsym}
\usepackage{subfigure}
\usepackage{pstricks}
\usepackage{color}

\usepackage{mathrsfs}
\usepackage{hyperref}

\newcommand{\be}{\begin{equation}}
\newcommand{\ee}{\end{equation}}
\newcommand{\bea}{\begin{eqnarray}}
\newcommand{\eea}{\end{eqnarray}}
\newcommand{\bml}{\begin{subequations}}
\newcommand{\eml}{\end{subequations}}
\newcommand{\bfig}{\begin{figure}}
\newcommand{\efig}{\end{figure}}

\newcommand{\bg}{\beta}

\newcommand{\del}{\delta}

\newcommand{\lb}{\lambda}

\newcommand{\og}{\omega}

\newcommand{\slashed}{\hspace{-1.1ex}/}

\begin{document}

\title{A universal bound on Quantum Chaos from Random Matrix Theory
%Out-of-time ordered correlators in Cosmology
%Quantum randomness from Fokker Planck Equation in Cosmology
}

\author{{{Sayantan Choudhury}}$^{1}$
 and {{Arkaprava Mukherjee}}$^{2}$
}
\affiliation{$^1$Quantum Gravity and Unified Theory and Theoretical Cosmology Group, Max Planck Institute for Gravitational Physics (Albert Einstein Institute),
	   Am M$\ddot{u}$hlenberg 1,
	   14476 Potsdam-Golm, Germany
}
\affiliation{$^2$Department of Physical Sciences, Indian Institute of Science Education and Research Kolkata,
Mohanpur, West Bengal 741246, India
}

%\vspace{5ex}
%\date{\today}
\begin{abstract}
In this article, using the principles of Random Matrix Theory (RMT), we give a measure of quantum chaos by quantifying Spectral From Factor (SFF) appearing from the computation of two point Out of Time Order Correlation function (OTOC) expressed in terms of square of the commutator bracket of quantum operators which are separated in time. We also provide a strict model independent bound on the measure of quantum chaos,  $-1/N(1-1/\pi)\leq {\bf SFF}\leq 0$ and $0\leq {\bf SFF}\leq 1/\pi N$, valid for thermal systems with large and small number of degrees of freedom respectively.  Based on the appropriate physical arguments we give a precise mathematical derivation to establish this alternative strict bound of quantum chaos.

\end{abstract}

%\pacs{98.80.-k ; 98.80.Cq ; 04.50.-h}

\maketitle
Quantum description of chaos has three important properties boundedness, exponential sensitivity and infinite recurrence. In the context of dynamical systems the concept of quantum chaos describes quantum signatures of classically chaotic systems. Quantum chaos can be formulated for two observable represented by hermitian operators $X(t)$ and $Y(t)$ using their commutator relation. This actually explains the perturbation effect of one operator $Y(t)$ on the measurement of other operator $X(t)$. Strength of this perturbation can be measured formulating a function:
\be C(t)=-Z^{-1}{\rm Tr}[e^{-\beta H}[X(t),Y(0)]^{2}]\ee at temperature $\bg=1/T$, also $Z$ is the partition function of the system and $H$ representing the system Hamiltonian. Here we assume that $X$ and $Y$ have zero one point function. Hence we use two point function for measuring quantum chaos, which technically has been explained by the late time behavior of the function $C(t)$ from which we can get a bound on quantum chaos. 

It has been previously shown that due to quantum effects two point function for chaos
decrease to a particular constant value. This decrease rate has an exponential growth with a factor $\lambda_{L}$ ({Lyapunov Exponent) which entirely depends on system properties and observable. Thus a bound on Lyapunov Exponent can be treated as measure of quantum chaos. Using quantum field theory it has been shown that an universal bound \cite{Maldacena:2015waa} on the {\it Lyapunov exponent} exists:
\be \lambda_L \leq 2\pi/\beta.\ee 
This bound is unique feature for all classes of out of equilibrium quantum field theory set up. 
We have also discussed the saturation of chaos bound at late time scale using Random matrix theory (RMT) which was previously discussed in the quantization of classical chaotic systems, usually in the semi-classical or high quantum-number regimes. For this discussion we construct Spectral From Factor (SFF)\cite{1997PhRvE..55.4067B}, which is arising from two point out of time orderd correlation (OTOC) function in RMT has been used to derive an alternative but strong bound on chaos. This gives us extra freedom to generalize our discussion for any quantum system with random interaction. This interaction has been included by a polynomial potential function of general order. Discussing the late time behavior of SFF. we get its upper bound. Also this approach is unique as the bound is valid for any arbitrary (infinite and finite) temperature. We also obtained a lower bound for SFF which depicts the minimal chaos a quantum system can have.  For arbitrary interaction where interaction potential is unknown we can get the eigen value distribution from RMT. For simplicity the arbitrary interaction potential can be expressed as a polynomial of different order.  From that eigen value distribution one can compute the partition function $Z$ in the present context. For our computation and prediction from RMT we have used any arbitrary order of  polynomial potential as interaction. Quantifying chaos using RMT is very useful when specific interaction at the level of action is not exactly known. To use RMT in this context we start with creating an statistical ensemble, that includes all possible interaction between energy levels of many-body Hamiltonian by various matrices in the ensemble. If the Hamiltonian is time-reversal then the symmetric distribution will be invariant under orthogonal transformation. In the thermodynamic limit ($N \rightarrow \infty$) eigen value distribution of random matrices showed a universal behavior , characterized by {\it Wigner's Semicircle law}. The results seemed to be applicable to a wide varied class of quantum systems displaying chaotic behavior. Even for many-body Hamiltonian (where $N$ finite) chaos can be treated, though it is better devised in systems where nearest neighbor spacing distribution (NNSD) of eigenvalues of the system is chaotic. For {\it Wigner Dyson} type distribution we get:
 \bea\label{g1}
P(\og = E_{n+1}-E_{n})=A_{\Gamma} \og^{\Gamma} e^{-\Gamma \og},
\eea
 where $\Gamma$ is a constant and for different ensemble it has particular values . SFF characterize spectrum of quantum system ( i.e. discreteness of energy spectrum) defined by :
 \bea{\bf SFF}=|Z(\bg + i t)|^{2}=\sum_{m,n} e^{-\bg (E_{m}+E_{n})} e^{-it(E_{m}-E_{n})}.
\eea
 Here it is important to note that, at infinite temperature, it pick out contribution form the difference between nearest neighbor energy eigenvalues at very late times. Averaging over Gaussian random matrices, SFF shows very particular behavior at large $N$, particularly initial decay followed by a linear rise and then a saturation after critical point. This saturation can be related to saturation limit for large $N$ as {\it bound on quantum chaos}.

In this context we use Gaussian matrix ensemble which is a collection of large number of matrices and these are filled with random numbers picked arbitrarily from a Gaussian probability distribution. Joint probability distribution of such random matrix can be expressed as, 
$P(M)dM=\exp\left(-\frac{1}{2}tr M^{2}\right)dM,$
where $N$ represent the rank of the random matrix. For any statistical ensemble of random matrices keeping this Haar measure invariant under similarity transformation, we get a constraint,
 $P(U^{-1}MU)=P(M)$. Here $U$ being an orthogonal or unitary matrix. Further integrating over the random matrix, partition function can be obtained for the Gaussian matrix ensemble by:
 %\begin{widetext}
 \bea \begin{array}{lll}
\displaystyle
Z=\int dM~ e^{-{\rm Tr}[V(M)]}=\prod_{i=1}^{N}\int d \lb_{i}~e^{-N^{2}S(\lb_{1},....\lb_{N})}                                                                                                                                                                                                                                                                                                                                                                                                                                                  
\end{array}~~~\eea
%\end{widetext}
where $V(M)$ denotes the random general polynomial potential. For generalized statistical ensemble expressing the problem in terms of the basis of eigenvalues of the matrix the partition function has action $S(\lb_{i})$ incorporated in it, which is given by:
\bea S(\lb_{1},....\lb_{N})=\frac{1}{N}\sum_{i=1}^{N} V(\lb_{i})+\bg \sum_{i< j}^{N}\log|\lb_{i}-\lb_{j}|.
 \eea                                                                                                                                         
 Here $\bg$ is fixed and eigen values are scaled by factor $\sqrt{N}$.
To find a solution we nee to extremise the action w.r.t  $\lb_{i}$ and use
 the method of resolvent to derive the expression for the partition function ($Z$) in the present context. The solution obtained in large $N$ limit can be compared with the solution obtained using WKB approximation from Schr{\"o}dinger equation. 

Just by knowing the general polynomial structure of the potential $V(x)$ one can able to find out the solution for the distribution of $\overline\og(x)=\lim_{N\to\infty}\og(x)$. {\it Wigner's semicircle law} predicts the probability density function of eigen values of many random matrices is a semi-circle as $N \rightarrow \infty$. Most general solution for the density function is :
  \bea \begin{array}{lll}
\displaystyle
\rho(\lb)=\frac{1}{2\pi}\sqrt{-\prod_{i=1}^{n}(\lb-d_{i-1})(\lb-d_{i})}\sum _{k=1}^{\infty} d_{n-k} \lambda ^{(n-k+1)},
%\\
%\displaystyle
 %M(\lb)=\sum _{k=1}^{\infty} a_{n-k} \lambda ^{2 (n-k)}\\
%\displaystyle
%\sigma(\lb)=\prod_{i=1}^{n}(\lb-a_{2i-1})(\lb-a_{2i}),
\end{array}~~~~ \eea
which satisfy normalization condition $\int_{{\rm supp }~\lb}d \lb~ \rho(\lb) =1$. Here we consider $n$ number of intervals on which $\rho(\lb)$ is supported and $d_{2i-1}$ and $da_{2i}$ are the end point of each interval. For general case the polynomial structure of interaction random potential $V(M)$, characterized in terms of the random matrix $M$ we can solve for its eigen value density.
 \bea V(M)=\sum _{i=1}^{\infty} C_{i} M^{i}
 \eea
 
Now we choose $\beta=0$ as quantum operator insertions around a thermal circular path is very straightforward at this limit.
Let us consider a quantum mechanical Hamiltonian operator $H$ operating on an ${\cal I}=2^n$ dimensional Hilbert space and consists of $n$ number of qbits. The averaged two point correlation function $\langle {\cal O}(0){\cal O}^{\dagger}(\tau)\rangle$ can be defined as:
%\begin{widetext}
 \bea \label{mnq}
\int d{\cal O}\langle {\cal O}(0){\cal O}^{\dagger}(\tau)\rangle &\equiv & \frac{1}{{\cal I}}\int d{\cal O}{\rm Tr}\left({\cal O}(e^{-iH\tau}){\cal O}^{\dagger}(e^{iH\tau})\right)\nonumber\\
\displaystyle
%~~~~~~~~~~~~~~~~~~~
&=&\frac{1}{{\cal I}^3}\sum^{{\cal I}^2}_{k=1}{\rm Tr}\left({\cal O}_k(e^{-H\tau}){\cal O}^{\dagger}_{k}(e^{iH\tau})\right)~~~~~~~%\\
\eea
%\end{widetext}
 Here ${\cal O}$ is the unitary operator integrated over a Haar measure on ${\cal U}(2^n)$. Here ${\cal O}_k$ represents Pauli operators which are  ${\cal I}^2=2^{2n}=4^n$ in total number.
First moment of the Haar ensemble, which is defined as:
\bea \begin{array}{lll}
\displaystyle
\int d{\cal O}~{\cal O}{\cal D}{\cal O}^{\dagger}=\frac{1}{{\cal I}}{\rm Tr}({\cal D})~{\bf I}
,~~ \int d{\cal O}~{\cal O}^{k}_{m}{\cal O}^{l}_{n}=\frac{1}{{\cal I}}\delta^{k}_{n}\delta^{l}_{m}.
\end{array}~~~~~~
\eea
From this we get the quantum averaged OTOC as:
 \bea \label{xx} 
\begin{array}{lll}
\displaystyle
\int d{\cal O}\langle {\cal O}(0){\cal O}^{\dagger}(\tau)\rangle\equiv \frac{1}{{\cal I}^2}|{\rm Tr}(\exp[-iHt])|^2 =\frac{1}{{\cal I}^2}{\bf SFF}(t)
%\nonumber \\&\propto& \textcolor{red}{\bf 2 p SFF}.,
\end{array}~~~~~
\eea 
For finite temperature we choose energy eigenvalue representation of OTOC :
\bea\begin{array}{lll}
\displaystyle
 {\cal C}(t)=\frac{1}{|Z(\beta)|^2}\sum_{n,m}c_{n,m}(t)e^{-\beta (E_n+E_m)}=\frac{|Z(\beta+it)|^2}{|Z(\beta)|^2}
\end{array}~~~~~
\eea
where the time dependent coefficient $c_{n,m}(t)$ is given by:
\bea\begin{array}{lll}
\displaystyle
\displaystyle
c_{n,m}(t)=-\langle n|[e^{-iHt},x]^2|m\rangle=\exp\left[-i(E_n-E_m)t\right].
\end{array}~~~~~
\eea 

Further consider a Thermo-field Double State (TDS) associated with quantum state at finite temperature. The time evolution of the TDS can be written as:
  \be |\Psi(\beta,t)\rangle=[Z(\beta)]^{-1/2}\sum_{n}e^{-\frac{\beta}{2}H}e^{iH\tau}. \ee 
from which SFF can be defined as:
\bea \begin{array}{lll}
\displaystyle
{\bf SFF}=|\langle \Psi(\beta,0|\Psi(\beta,t)\rangle|^2 =\frac{|Z(\bg + i t)|^{2}}{|Z(\beta)|^2}= {\cal C}(t).~~~
\end{array}
 \eea
Additionally, in this context, SFF helps in understanding the time dynamics as well as to analyze the discreteness of the spectrum.
To quantify SFF we define a two point (Green's) function $G(\bg,t)$ as:
\begin{widetext}
\bea
%~~~~~~~~~~~~~~
 &&G(\bg,t)=\frac{\langle|Z(\bg+it)|^{2}\rangle_{\rm GUE}}{\langle Z(\bg)\rangle^{2}_{\rm GUE}}%\\
%\\
\displaystyle
=\frac{\int_{{\rm supp}~ \overline\rho} d\lb~ d\mu~ e^{-\bg(\lb+\mu)}~e^{-it(\lb-\mu)}\langle D(\lb)D(\mu)\rangle_{\rm GUE}}{\int_{{\rm supp }~ \overline\rho} d\lb~ d\mu~ e^{-\bg(\lb+\mu)}\langle D(\lb)\rangle_{\rm GUE} \langle D(\mu)\rangle_{\rm GUE}}=G_{dc}(\beta,t)+G_{c}(\beta,t).
\\
&& G_{dc}(\beta,t)=\left[\frac{\langle Z(\bg+it)\rangle_{\rm GUE} \langle Z(\bg-it)\rangle_{\rm GUE}}{\langle Z(\bg)\rangle^{2}_{\rm GUE}}\right], ~~~G_{c}(\beta,t)=\frac{\int d\lb~ d\mu~ e^{-it(\lb-\mu)}~\langle \del D(\lb) \del D(\mu)\rangle_{\rm GUE}}{\int d\lb~ d\mu~ \langle D(\lb)\rangle_{\rm GUE} \langle D(\mu)\rangle_{\rm GUE}}.
\eea
 \end{widetext}
 Here, $D(\lb)=\overline\rho(\lb)=\lim_{N\to\infty}\rho(\lb)$ represents the eigen value density at large $N$. Here we use the fact,
 $\langle D(\lb)D(\mu)\rangle_{\bf c}= \langle \del D(\lb) \del D(\mu)\rangle $.
% the mean level density can be normalized in a semi circle using the following two conditions:
 %\bea \int_{-2a}^{2a} d\lb~D(\lb) &=&N,\\
%\int_{-2a}^{2a} d\lb~\rho(\lb) &=&1.\eea

In this context the partition function is computed for general order polynomial as:
\begin{widetext}
\bea 
%\displaystyle\langle Z(\bg\pm it)\rangle_{\rm GUE}=\frac{1}{\pi}\int_{-2a}^{2a} d \lb~ e^{\mp it\lb}~e^{-\beta\lambda}~ \sqrt{4 a^2-\lambda ^2}\times\sum _{k=1}^n d_{n-k} \lambda ^{2 (n-k)}~~~~~~\forall~ {\rm even~n}\\
%\displaystyle
%\equiv\sum^{n}_{k=1}d_{n-k}~ a^2 (-a^2)^{-2 k} 4^{n-k}[\{(e^{2 i \pi  k}+e^{2 i \pi  n}) a^{2 (k+n)} 
%\Gamma(-k+n+\frac{1}{2})
%\, _1\tilde{F}_2(-k+n+\frac{1}{2};\frac{1}{2},-k+n+2;a^2 (\beta\pm i t)^2)\}\\
%\displaystyle
%~~~~+\{a (\beta\pm i t)(a^{2 k} (-a)^{2 n}-(-a)^{2 k} a^{2 n}) \Gamma (-k+n+1)
%\, _1\tilde{F}_2(-k+n+1;\frac{3}{2},-k+n+\frac{5}{2};a^2 (\beta\pm i t)^2)]\}~~~~~~\forall~ {\rm even~n} 
&&\langle Z(\bg\pm it)\rangle_{\rm GUE}=\frac{1}{\pi}\int_{-2a}^{2a} d \lb~ e^{\mp it\lb}~e^{-\beta\lambda}~ \sqrt{4 a^2-\lambda ^2}\times\sum _{k=1}^n d_{n-k} \lambda ^{-k+n+1}\nonumber\\
&&\equiv \sum^{n}_{k=1} d_{n-k}(-1)^{1-k}a^{3-2k} 2^{1-k+n}\left[\left\{a^k (-a)^n-(-a)^k a^n\right\} \Gamma \left(\frac{1}{2} (-k+n+2)\right) \, _1\tilde{F}_2\left(\frac{1}{2} (-k+n+2);\frac{1}{2},\frac{1}{2} (-k+n+5);-a^2 t ^2\right) \right.\nonumber\\ &&\left.
~~~~~~~~+i a t \left\{(-a)^k a^n+a^k (-a)^n\right\} \Gamma \left(\frac{1}{2} (-k+n+3)\right) \, _1\tilde{F}_2\left(\frac{1}{2} (-k+n+3);\frac{3}{2},\frac{1}{2} (-k+n+6);-a^2 t ^2\right)\right]
\eea
\end{widetext}

For different polynomial potentials calculating the expansion coefficients $d_{n-k}$ from method described earlier one can get the exact form of $Z(\bg\pm it)$. On the other hand, the 
two point function arising from the connected part can be expressed as:
 \bea\label{xc11}
\langle \del D(\lb) \del D(\mu)\rangle=-\frac{\sin^{2}[N(\lb-\mu)]}{(\pi N(\lb-\mu))^{2}}+\frac{1}{\pi N}\del(\lb-\mu),~~~~
\eea
where  $1/N^{2}$ part with sine squared function gives the ramp and $1/N$ part with Delta function gives the plateau region which dominates over the other part. To perform the integral appearing in the expression for connected part of the Green's function we further substitute,  
$\lb+\mu=E,
\lb-\mu=\og.$ Since the integral over $E$ gives Dirac Delta function we choose $E=0$ in which perform the computation. Consequently we get:
  \bea\begin{array}{lll}\label{cvcvx}
		\displaystyle   S(t)=N^2G_c(\tau)=\left\{\begin{array}{lll}
			\displaystyle  
			\frac{t}{4\pi^2 N^2}-\frac{1}{N}+\frac{1}{\pi N}\,,~ &
			\mbox{\small  \textcolor{black}{\bf  {$\tau<2\pi N$ }}}  \\ \\
			\displaystyle  
			\frac{1}{\pi N}.~&
			\mbox{\small  \textcolor{black}{\bf  {$\tau>2\pi N$}}}  
		\end{array}
		\right.
	\end{array},\eea
From the obtained result it is clearly observed that we get the linear growth in the region $\tau<2\pi N$ and the constant plateau type behaviour in the region $\tau>2\pi N$. Also the change in behavior of the SFF at the transition point  is abrupt in nature.

 To derive the bound on SFF  we first use the asymptotic behavior of regularized HypergeomtericPFQ function \cite{Bound}, which is given below:
\be
\lim_{t\to\infty} \, _1\tilde{F}_2\left[A(k);B(k),C(k);a^2 (\beta\pm  i t)^2)\right] =0~~\forall k=1,\cdots,n
\ee
where $A(k),B(k)$ and $C(k)$ are time independent variables. Also $n$ represents the highest order of the polynomial. Consequently, the asymptotic behavior of the connected and disconnected part of the Green's function in the regime $t> 2 \pi N$ can be expressed with finite $N$ as:
\bea\begin{array}{lll}\label{cbnb}
		\displaystyle   \lim_{t\to\infty} G_c(\beta,t)=\left\{\begin{array}{lll}
			\displaystyle  
			\frac{1}{\pi N}\,,~ &
			\mbox{\small  \textcolor{black}{ {$\beta\neq 0$ }}}  \\ \\
			\displaystyle  
			0.~&
			\mbox{\small  \textcolor{black}{{$\beta=0$}}}  
		\end{array}
		\right.
	\end{array},\\ \label{limit1}
\lim_{t\to\infty} G_{dc}(\beta,t)=0~~~~\textcolor{black}{\forall~  t>2\pi N},~ \forall \beta.\eea
Finally, adding the contribution from the disconnected and connected part of the Green's function in the asymptotic limit we get the following simplified expression for SFF in the regime $t> 2 \pi N$ at finite $N$:
\bea\begin{array}{lll}\label{bnbn}
		&&\displaystyle  \lim_{t\to\infty}{\bf SFF}(\beta,t>2 \pi N)=\left\{\begin{array}{lll}
			\displaystyle  
			\frac{1}{\pi N}\,,~ &
			\mbox{\small  \textcolor{black}{ {$\beta\neq 0$ }}}  \\ \\
			\displaystyle  
			0.~&
			\mbox{\small  \textcolor{black}{{$\beta=0$}}}  
		\end{array}
		\right.
	\end{array},\eea
For $t\to \infty$ asymptotic limit till now we have only considered the part of SFF appearing only after $t>2\pi N$ with finite $N$. On the other hand the main obstacle of 
taking $t\to \infty$ asymptotic limit in the $t<2\pi N$ region with finite $N$ is  we get divergent contribution in the connected part of the Green's function $G_c$ from the term $t/(2\pi N)^{2}\to \infty$. Also the
disconnected part of the Green's function behave same as Eq.~(\ref{limit1}). Consequently in the regime $t<2\pi N$ with finite $N$ the SFF can be computed as:
\be  \lim_{t\to\infty}{\bf SFF}(\beta,t< 2 \pi N)=\infty~~~~\forall~ \beta.\ee
Hence combining both the contribution from connected and disconnected part of the total Green's function we get the following upper and lower bound on SFF in the regime $t>2\pi N$, as given by:
\bea \begin{array}{lll}\label{e1}
\displaystyle  
0\leq {\bf SFF}(\beta,t>2\pi N)\leq \frac{1}{\pi N}~~\forall~ \textcolor{black}{~~0\leq \beta\leq \infty}~.
\end{array}
\eea

With large $N$ asymptotic limit in the $t<2\pi N$ region gives finite contribution to the connected and disconnected part of the Green's function as given by:
\bea\begin{array}{lll}
\displaystyle
 \lim_{t\to\infty} G_{c}(\beta,t)\simeq-\frac{1}{N}\left(1-\frac{1}{\pi }\right),
%~~~~\forall~\beta
%\\
\displaystyle
\lim_{t\to\infty} G_{dc}(\beta,t)=0~\forall \beta.~~~~~
\end{array} 
\eea
Further adding both the contribution from connected and disconnected part of the Green's function for the asymptotic region $t<2 \pi N$ with large $N$ we get the following upper and lower bound on SFF, as given by:
\bea  \label{d1} 
\begin{array}{lll}
\displaystyle
-\frac{1}{N}\left(1-\frac{1}{\pi }\right)\leq {\bf SFF}(\beta,t<2 \pi N)\leq 0 
~~\forall~~0\leq \beta\leq \infty.
\end{array}~~~~~
\eea
\begin{figure}[htb]
\centering
{
    \includegraphics[width=8.5cm,height=8.5cm] {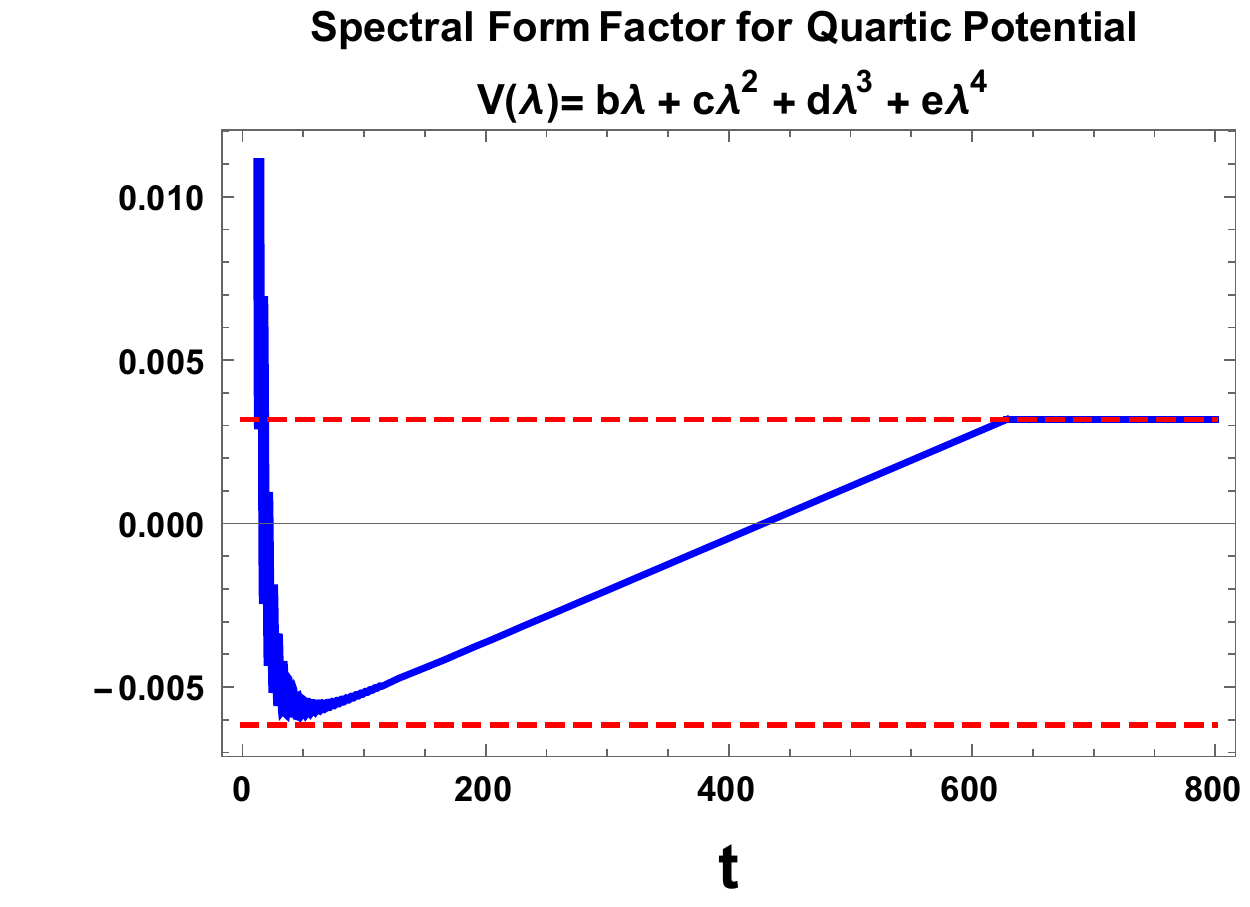}
    \label{SFFbound2}
}
\end{figure}
\begin{figure}[htb]
	\centering  
{    \includegraphics[width=8.5cm,height=8.5cm] {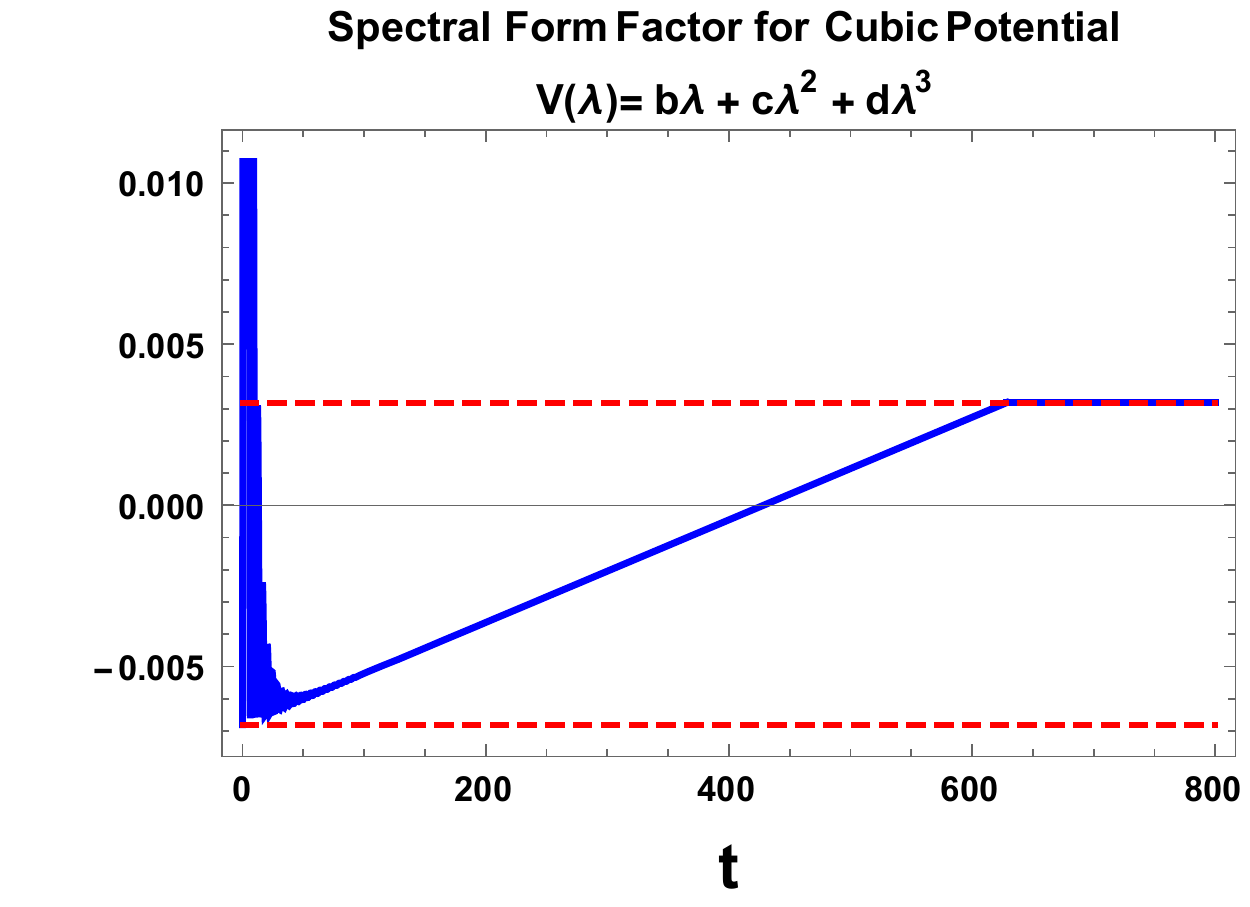}
    \label{SFFbound4}
}

\caption{Behavior of SFF at finite temperature for different polynomial potentials.}
\label{sfd}
\end{figure} 
In figure.~(\ref{sfd}) we have shown the time dependent behavior of SFF at fixed finite temperatures. Additionally, we have explicitly depicted the saturation bound on quantum chaos both in $t>2 \pi N$ (for finite $N$) and $t< 2 \pi N$ (for large $N$) region.
For any general polynomial potential explained method can be applied and bound can be obtained. The symmetry property for $\lambda \rightarrow -\lambda$ is lost and non-vanishing support should be chosen \cite{brezin1978}. Particularly for odd polynomial potential the resolvent method is not applicable. But for the even and general one it works perfectly well.

 From our discussion using RMT we have derived a strict bound on SFF for both large $N$ and finite $N$ situations, which ultimately put stringent constraint on quantum chaos for randomly interacting system. Using this approach we find that our predicted bound for quantum chaos from RMT is independent on temperature, which further implies that the derived result is universal in nature. Moreover, using the lower bound on SFF one can precisely comment on the minimal chaotic nature of a random system from RMT. Also, this obtained result can be further used to describe out of equilibrium aspects in cosmology~\cite{Choudhury:2018rjl}, specifically to quantify reheating temperature and also in the context of quantum theory of black-hole~\cite{Shenker:2013pqa,Cotler:2016fpe,Hayden:2007cs,Sekino:2008he,Shenker:2013pqa}.
 
%%%%%%%%%%%%%%%%%%%%%%%%%%%%%%%%%%%%%%%%%%%%%%%%%%%%%%%%%%%%%%%%%%%%%%%%%%%%%%%%
{\bf Acknowledgement:}~SC would like to thank Max Planck Institute for Gravitational Physics, Albert Einstein Institute for providing the Post-Doctoral
	Research Fellowship.

${}^{1}$sayantan.choudhury@aei.mpg.de,\\
$~~~~{}^{2}$arka262016@gmail.com
%%%%%%%%%%%%%%%%%%%%%%%%%%%%%%%%%%%%%%%%%%%%%%%%%%%%%%%%%%%%%%%%%%%%%%%%%%%%%%%%%%%%%%%%%%%%%%%%%%%%%%%%%%%%%%%%%%%%%%%%%%%
%%%%%%%%%%%%%%%%%%%%%%%%%%%%%%%%%%%%%%%%%%%%%%%%%%%%%%%%%%%%%%%%%%%%%%%%%%%%%%%%%%%%%%%%%%%%%%%%%%%%%%%%%%%%%%%%%%%%%%%%%%%%%%%%%%%%%%%%%%%%%%%%%%%%%%%%%%%%%%%%%%%%%%%%%%%%%%%%%%%%%%%%%%%%%%%%%%%%%%%%%%%%%%%%%%%%%%%%%%%

%%%%%%%%%%%%%%%%%%%%%%%%%%%%%%%%%%%%%%%%%%%%%%%%%%%%%%%%%%%%%%%%%%%%%%%%%%%%%%%%%%%%%%%%%%%%%%%%%%%%%%%%%%%%%%%%%%%%%%%%%%%%%%%%%%%%%%%%%%%%%%%%%%%%%%%%%%%%%%%%%%%%%%%%%%%%%%%%%%%%%%%%%%%
%%%%%%%%%%%%%%%%%%%%%%%%%%%%%%%%%%%%%%%%%%%%%%%%%%%%%%%%%%%%%%%%%%%%%%%%%%%%%%%%%%%%%%%%%%%%%%%%%%%%%%%%%%%%%%%%%%%%%%%%%%%%%%%%%%%%%%%%%%%%%%%%%%%%%%%%%%%%%%%%%%%%%%%%%%%

%\end{thebibliography}


\begin{references}

\bibitem{Maldacena:2015waa}
J.~Maldacena, S.~H.~Shenker and D.~Stanford,
%``A bound on chaos,''
JHEP {1608} (2016) 106


\bibitem{1997PhRvE..55.4067B}	
E.Brezin, S. Hikami, 
Phys. Rev. E, 55, 4 (1997) 4067.


\bibitem{Bound}
Cvitković, Mislav; Smith, Ana-Sunčana; Pande, Jayant
Journal of Phys. A,  50, 26  (2017) 265206.
	
\bibitem{brezin1978}
Brézin, E.; Itzykson, C.; Parisi, G.; Zuber, J. B. Planar diagrams. Comm. Math. Phys. 59, 1 (1978) 35. 






%\cite{Choudhury:2018rjl}
\bibitem{Choudhury:2018rjl}
S.~Choudhury, A.~Mukherjee, P.~Chauhan and S.~Bhattacherjee,
%``Quantum Out-of-Equilibrium Cosmology,''
arXiv: 1809.02732 [hep-th].
%%CITATION = ARXIV:1809.02732;%%

\bibitem{Cotler:2016fpe}
J.~S.~Cotler {\it et al.},
%``Black Holes and Random Matrices,''
JHEP {\bf 1705} (2017) 118


%\cite{Hayden:2007cs}
\bibitem{Hayden:2007cs}
P.~Hayden and J.~Preskill,
%``Black holes as mirrors: Quantum information in random subsystems,''
JHEP {\bf 0709} (2007) 120

\bibitem{Sekino:2008he}
Y.~Sekino and L.~Susskind,
%``Fast Scramblers,''
JHEP {\bf 0810} (2008) 065

\bibitem{Shenker:2013pqa}
S.~H.~Shenker and D.~Stanford,
%``Black holes and the butterfly effect,''
JHEP {\bf 1403} (2014) 067

\end{references}
\end{document}